\documentstyle [prl,aps,epsfig]{revtex}

\newcommand{\la}[1]{\label{#1}}

\newcommand{\be}{\begin{equation}}
\newcommand{\ee}{\end{equation}}
\newcommand{\ba}{\begin{array}}
\newcommand{\ea}{\end{array}}
\newcommand{\bastar}{\begin{eqnarray*}}
\newcommand{\eastar}{\end{eqnarray*}}

\newcommand{\G}{\Gamma}
\newcommand{\GR}{\Gamma^{(0)}_{R}[\hat\phi]}
\newcommand{\GNR}{\Gamma^{(0)}_{NR}[\hat\phi,c,\bar{c};\alpha ]}
\newcommand{\f}{{\phi}}                
\newcommand{\m}{{\mu}}
\newcommand{\al}{{\alpha}}
\newcommand{\intg}{{\int d^{4}x\,}}
\newcommand{\lam}{{\lambda}}
\newcommand{\fh}{{\hat\phi}}
\newcommand{\pa}{\partial}
\newcommand{\es}{\\[3mm]}
\newcommand{\pad}[2]{{\frac{\partial #1}{\partial #2}}}
\newcommand{\fud}[2]  {{\displaystyle{\frac{\delta #1}{\delta #2}}}}

\newskip\humongous \humongous=0pt plus 1000pt minus 1000pt

\newif\ifdtup

\relax

\begin{document}
%
\title  {Renormalizability of Nonrenormalizable Field Theories}
\bigskip

\author{Alberto Blasi$^{* }$, Nicola Maggiore$^{*}$,
Silvio P.Sorella$^{** }$ and Luiz C.Q.Vilar$^{**}$} 

\address{
$^*$Dipartimento di Fisica, Universit\`a di Genova, via Dodecaneso 
33,I-16146 Genova,Italy\\
$^{**}$UERJ, Universidade do Estado do Rio de Janeiro,
Departamento de F\'{\i}sica Te\'{o}rica, 
Rua S\~ao Francisco Xavier, 524,
20550-013, Maracan\~{a}, Rio de Janeiro, Brazil\\
{\scriptsize \bf BLASI@GE.GE.INFN.IT, 
MAGGIORE@GE.INFN.IT,SORELLA@DFT.IF.UERJ.BR,LCQVILAR@DFT.IF.UERJ.BR} 
\\ \vskip 0.3cm
}

\maketitle

\begin{abstract}
We give a simple and elegant proof of the Equivalence Theorem, stating 
that two field theories related by nonlinear field transformations have 
the same S matrix. We are thus able to identify a subclass of 
nonrenormalizable field theories which are actually physically 
equivalent to renormalizable ones. Our strategy is to show by means 
of the BRS formalism that the 
``nonrenormalizable'' part of such fake nonrenormalizable theories, is 
a kind of gauge fixing, being confined in the cohomologically trivial 
sector of the theory. 
\end{abstract}

\widetext
\bigskip

Recently there has been a renewed interest, triggered by the work of 
Gomis and Weinberg~\cite{gw} on the apparently nonrenormalizable 
theories. The main point was analyzed by Berg\`ere and Lam~\cite{bl} 
who showed that two quantum field theories  related by a 
nonlinear field transformation of the kind
\be
\f = \fh + \al\,\fh^{2}g(\fh;\al)
\la{fh}
\ee
have the same S-matrix. This statement is known in the literature as 
the ``equivalence theorem'' since more than fifty years~\cite{thm}, and we
propose here an alternative 
very simple proof, which is easily adaptable to all situations, 
since it relies neither on the use of the equations of motion, nor on 
any particular renormalization scheme.

The strategy is to approach 
the problem with the technology of nilpotent operators, as it is 
applied in gauge field theories, and hence to interpret the effect 
on the  action generated by the nonlinear part of the transformation~(\ref{fh}) 
as a ``gauge fixing'' term.
If this is possible, we have immediately at our disposal the standard 
results of gauge field theories which insure that ``physics'' is 
independent from the gauge choice. 

In order to emphasize the relevant features we shall treat only the 
simplest case, $i.e.$ a scalar field theory with quartic interaction 
in four dimensions; the method can be straightforwardly extended to 
the physically interesting cases.

We begin with the renormalizable classical action
\be
\G_{R}^{(0)}[\f]\ =\ \intg \left ( \frac{1}{2}\pa_{\m}\f\pa_{\m}\f + 
\frac{1}{2} m^{2}\f^{2} + \frac{1}{4!}\lam\f^{4} \right )
\la{gammaR}
\ee
and the related path integral representation for the vertex functional
\be
\G_{R}\ =\ \int D(\f) \exp{(-\G_{R}^{(0)}[\f])}
\ee
Now we perform the nonlinear field redefinition (\ref{fh}), where the 
nonlinear part of the field transformation is identified by the 
parameter $\al$, and $g(\fh;\al)$ is an analytic function of both $\f$ 
and $\al$. The introduction of the $\al$-parameter, although it could appear 
as a computational artifact, is indeed natural since to preserve the 
dimensionally homogeneous character of (\ref{fh}), $\al$ has the 
dimension of $[\frac{1}{m}]^{d_{\f}}$.
By applying (\ref{fh}) to (\ref{gammaR}), we obtain a new classical 
action, but we have also to take into account the Jacobian of the 
field transformation, which is conveniently exponentiated by means of 
anticommuting variables $\bar{c(x)}$ and $c(x)$. Thus we find a new 
classical action, non renormalizable by power counting
\be
\GNR \ =\ \GR + \al\G^{(1)}[\fh;\al ] 
+ \intg  \bar{c}\left (1 + 2\al\fh 
g(\fh;\al) + \al\fh^{2}g'(\fh;\al)\right ) c
\la{gammaNR}
\ee
where $\al\G^{(1)}[\fh;\al ]$ is obtained from 
\be
 \al\G^{(1)}[\fh;\al ]={\Gamma^{(0)}_{R}[\fh + \al\,\fh^{2}g(\fh;\al)]} - {\Gamma^{(0)}_{R}[\hat\phi]} 
\la{gamma1}
\ee
The corresponding path integral formulation for the proper 
functional now reads
\be
\G_{NR}\ =\ \int D(\fh)D(c)D(\bar{c})\exp{(-\GNR)}
\ee
It is precisely the part of $\GNR$ not coinciding with 
$\GR$, which we would like 
to identify as a ``gauge fixing term'' with ``gauge parameter'' $\al$. 

With this in mind we introduce two ghosts $\beta$ and $b(x)$ of which 
the first is global, while the second is local, and the BRS 
transformations
\be\ba{l}
s \fh(x) = s c(x) = s \bar{c}(x) = s\al=0 \ ,\es

s b(x)   = \bar{c}(x)\ ,\es
s \beta    = \al
\ea\la{brs}\ee
Correspondingly, the classical action (\ref{gammaNR}) reads
\be
\GNR = \GR + s\, Y[\fh,c,\bar{c};\al]
\ee
where
\be
Y[\fh,c,\bar{c};\al]=
\beta \G^{(1)} + \intg b (1 + 2\al\fh 
g(\fh;\al) + \al\fh^{2}g'(\fh;\al) ) c
\ee
The classical action $\GNR$ satisfies the linear Slavnov--Taylor 
identity
\be
{\cal S}\GNR = \left (
\intg \bar{c}(x)\fud{}{b(x)} + \al\pad{}{\beta}
\right)\GNR = 0
\la{slavnov}\ee
Moreover, the action is uncharged with respect to the Faddeev-Popov 
assignments written in the Table 
\begin{center}
\begin{tabular}{|l||r|r|r|r||r|r|}\hline
 & $\fh$ & $c$ & $\bar{c}$ & $b$ & $\al$ & $\beta$
\\ \hline
$\Phi\Pi$ & $0$ & $1$  & $-1$ & $-2$ & $0$ & $-1$
\\ \hline
\end{tabular}

\vspace{.2cm}{\footnotesize {\bf Table }
Faddeev--Popov charges.}
\end{center}

To make contact with the initial problem, and to identify the physical 
subspace of our example, we restrict the space to that of analytic 
functions of the $\al$-parameter. Within this subspace we can analyze 
the cohomology of the BRS operator (\ref{brs}) and easily find that 
it contains only $\al$-independent local functionals of $\fh(x)$ and 
$c(x)$, since $\{b(x);\bar{c}(x)\}$ and $\{\beta;\al\}$ appear in 
(\ref{brs}) as BRS-doublets~\cite{piguet-sorella}.

Thus we have the parametric equation
\be
\al\pad{}{\al}\G = s \left( \intg  \widehat{X} \right ) \cdot \G
\la{goal}\ee
for a suitable local functional $\widehat{X}$.
Notice that we have to employ the $\al\pad{}{\al}$ operator which 
leaves the cohomology invariant and not simply the $\pad{}{\al}$ 
operator which mixes the cohomological separation of the target space.
Indeed, from the expression~(\ref{slavnov}) we have that 
\be
\pad{}{\al}{\cal S}{\cal F} - {\cal S}\pad{}{\al}{\cal F}
=\pad{}{\beta}{\cal F}
\la{algebra}\ee
where ${\cal F}$ is a generic functional. Hence
\be
\pad{}{\al}\GNR 
= 
\pad{}{\al} {\cal S} Y[\fh,c,\bar{c};\al]
= 
\left({\cal S} \pad{}{\al}+\pad{}{\beta}\right) Y[\fh,c,\bar{c};\al]
=
{\cal S}\pad{}{\al}Y[\fh,c,\bar{c};\al] + \G^{(1)}
\ee
So $\pad{}{\al}\GNR$ is not cohomologically trivial, but $\al\pad{}{\al}\GNR$
on the contrary does, since
\be
\al\pad{}{\al}\GNR = {\cal S} \left(\al\pad{}{\al} Y[\fh,c,\bar{c};\al] 
+ \beta \G^{(1)} \right)
\ee
Equation (\ref{goal}) is the statement that only the 
$\al$-independent Green functions are ``physical'' and these are 
built with the vertices and the propagator obtained by $\GR$. One 
final remark may be in order; to implement the stability of the 
theory, we may impose on $\G$ the further 
conditions 
\be
\pad{\G}{\beta} = \fud{\G}{b(x)} = 0
\ee 
which are trivially true at the classical level.

Finally, we would like to provide a simple method to decide whether or 
not a theory specified by a classical action which appears to be 
nonrenormalizable by power counting can be obtained by a power 
counting renormalizable action through a nonlinear field 
redefinition. To be definite, consider the example treated in this 
letter, $i.e.$ a scalar field $\f(x)$ with a certain classical action 
$\G_{NR}(\f)$. First collect in $\G_{R}(\f)$ all terms which are power 
counting renormalizable. The remaining contributions, being non power 
counting renormalizable, contain at least a power of a parameter 
$\al$ with the dimension of an inverse of a mass; therefore we can 
write
\be
\G_{NR}(\f) = \G_{R}(\f) + \al \G^{*}(\f)
\ee
If $\G_{NR}(\f)$ can be obtained from $\G_{R}(\f)$ by a nonlinear 
field transformation
\be
\f\longrightarrow\f+\al X(\f;\al)
\ee
we have to solve for $X(\f;\al)$ the equation
\be
 \G_{R}(\f+\al X(\f;\al)) - \G_{R}(\f) = \al \G^{*}(\f)
\ee
A very simple criterion is to analyze it in a descending way, 
beginning from the highest order monomial in $\G^{*}(\f)$ and 
$\G_{R}(\f)$ and remembering that $X(\f;\al)$ is at least quadratic in 
$\f(x)$. For example, an action $\G_{NR}(\f)$ which contains $\f^{2}$, 
$\f^{4}$ and $\f^{6}$ terms only is truly non power counting 
renormalizable since, through a bilinear transformation containing 
only one parameter we are bound 
to obtain a $\f^{8}$ contribution, too.  

{\bf Acknowledgments}
We are grateful to the organizing Committee for the kind hospitality 
during the conference {\it Quantization, Generalized BRS Cohomology 
and Anomalies}, 
held at the Erwin Schroedinger International Institute (Vienna), 
from september 28 to october 7, 1998.

\vfill
                      
\eject


\begin{thebibliography}{99}


\bibitem{gw} J.Gomis and S.Weinberg, Nucl. Phys. { B469} 473 (1996); 
\bibitem{bl} M.C.Berg\`ere and Y.-M.P.Lam, Phys. Rev. { D13} 3247 (1976);
\bibitem{thm} F.J.Dyson, Phys. Rev. 73 829 (1948);
              R.Haag, Phys. Rev. 112 669 (1958);
              S.Kamefuchi, L.O'Raifertaigh and A.Salam, Nucl. Phys. 28 
              529 (1961);
              A.Salam and J.Strathdee, Phys. Rev. D2 2869 (1970);
              Y.-M.P.Lam, Phys. Rev. D7 2943 (1973);
\bibitem{piguet-sorella} O. Piguet and S.P. Sorella, ``Algebraic
                 Renormalization'', {\em Lecture Notes in Physics,
                 Vol.} m28
                 ({\em Springer Verlag, Berlin,
                 Heidelberg}, 1995).
\end{thebibliography}
\end{document}